\documentstyle[twoside,fleqn,espcrc2]{article}

\newcommand{\lsim}{\mbox{\raisebox{-.9ex}{~$\stackrel{\mbox{$<$}}{\sim}$~}}}

\title{Towards a model of quintessential inflation}

\author{K. Dimopoulos\address{Astroparticle and High Energy Physics group, 
        Instituto de Fisica Corpuscular,\\
	Universitat de Valencia/Consejo Superior de Investigaciones 
	Cient\'{\i}ficas,\\ 
	Edificio Institutos de Paterna, 
	Apartado de Correos 22085, 46071 Valencia, Spain}%
        \thanks{This work is done in collaboration with J.W.F. Valle and 
	has been supported by {\sc dgicyt} grant and the EU under the 
	contract {\sc erbfmrx-ct96-0090}.}}
       
\begin{document}

\begin{abstract}
A model of quintessential inflation is presented, which
manages to achieve the requirements of both inflation and quintessence
with natural values of the mass-scales and parameters.
\end{abstract}

\maketitle

\section{INTRODUCTION}

Recent observations suggest that the Universe is 
entering a phase of accelerated expansion \cite{SNIa}. This is also in 
agreement with other observational facts which favor a Universe dominated 
at present by a dark energy component \cite{dark}. Many consider 
that this required dark energy is provided by a non-vanishing cosmological 
constant, which, however, needs to be extremely fine-tuned \cite{fine}. The 
problem may be overcome if the necessary dark energy is provided by a scalar 
field, called quintessence, whose energy density comes to dominate the 
Universe today \cite{quint}. Unfortunately, some fine-tuning is still required 
for the parameters of the quintessential potential \cite{qprob}. Moreover, 
quintessence introduces several other problems, such as tuning of initial 
conditions.

A possible solution to the above puzzle is linking the quintessence 
paradigm with inflationary theory. This is a rather natural course of action
since both scenarios are based on the same idea, i.e. that the Universe 
undergoes accelerated expansion when dominated by the potential energy of a 
scalar field $\phi$, which rolls down its almost flat potential. In such 
models the quintessence field is identified with the inflaton. 

Quintessential inflation has many advantages. For a start, the 
introduction of yet another scalar field, whose nature and origin is 
unaccounted for, is avoided. Furthermore, a single theoretical framework 
can be used to accommodate both processes. Also, it is possible to
avoid introducing and tuning too many mass scales and model parameters. 
In that way one may solve, or at least link, the fine-tuning problems of 
quintessence with that of inflation. Finally, one dispenses with several
other problems such as the fine tuning of initial conditions, which for the 
quintessence part are determined by the end of inflation, whereas for 
inflation itself they can be simply chaotic.

Certain special requirements need to be satisfied in order to construct a 
successful quintessential inflationary model. Firstly, $V(\phi)$ should 
feature a ``quintessential tail'', 
\mbox{$V(\phi\!\rightarrow\!\infty)\rightarrow 0$}, so that $\phi$ has not 
reached the minimum yet. Secondly, $\phi$ should not be coupled to the 
Standard Model (SM) fields in order not to decay into SM particles at the end 
of inflation but, instead, survive until today. Such a ``sterile'' inflaton 
dispenses also with one typical fine tuning problem of inflation, mainly that 
the couplings of the inflaton to the SM should be large enough to ensure 
prompt reheating but also small enough not to destabilize the flatness of the 
slow-roll potential. In quintessential inflation reheating is achieved only 
through gravitational particle production.

Few models of quintessential inflation exist in the literature 
\cite{quinfkin}\cite{qinf}, 
because it is rather difficult to manage in a single stroke to solve the 
horizon, flatness, monopole and coincidence problems, while also providing 
the seeds for Large Scale Structure (LSS) formation and for the anisotropies 
of the Cosmic Microwave Background Radiation (CMBR). Our model manages to 
achieve all the above while using natural values for its mass-scales and 
parameters.

\section{THE MODEL}

Consider the potential \cite{mine},

\begin{equation}
V(\phi) =
M^4(\phi)\left[1+\cos\Big(\,
\frac{\pi\phi}{\sqrt{\phi^2+
\mu^2}}\,\Big)\right]
\label{V}
\end{equation}
and

\begin{equation}
M(\phi)\equiv M
\left\{\,\frac{M}{\mu}+\frac{\mu}{M}
\frac{\sqrt{z}\exp(-z\phi/m_P)}{\cosh(\lambda\phi/\mu)
\cos[\omega]}
\,\right\}
\label{Mf}
\end{equation}
where \mbox{$[\omega]\equiv$ {\large
$\Big(\frac{\omega\pi\phi}{\sqrt{\phi^2+\mu^2}}\Big)$}},
\mbox{$0<z,\lambda,\omega\lsim 1$}, \mbox{$M\ll\mu$} and
\mbox{$m_P\equiv M_P/\sqrt{8\pi}$} with $M_P$ being the Planck mass. 
In order for \mbox{$V(\phi\rightarrow\infty)\rightarrow 0$}, we
require \mbox{$\omega<\frac{1}{2}$}. From the above we have,

\begin{eqnarray}
 & \hspace{-0.5cm} & V'=-(V/\mu)\times\nonumber\\
 & \hspace{-0.5cm} & 
\left\{\,\mbox{\large $\frac{4\,[z_m+\lambda\tanh(\lambda\phi/\mu)]-
4\omega\pi\tan[\omega][1+(\phi/\mu)^2]^{-3/2}}{\frac{1}{\sqrt{z}}
(\frac{M}{\mu})
\cosh(\lambda\phi/\mu)\exp(z\phi/m_P)\cos[\omega]+1}$}\;+\right.\nonumber\\
 & \hspace{-0.5cm} & \left.\hspace{0.5cm}
+\;\pi\tan\Big(\,\mbox{\large $\frac{\pi\phi/2}{\sqrt{\phi^2+\mu^2}}$}\,\Big)
[1+(\phi/\mu)^2]^{-3/2}\right\}
\end{eqnarray}
where the prime denotes derivative with respect to $\phi$ and 
\mbox{$z_m\equiv z(\frac{\mu}{m_P})$}.

The equations of motion are,

\begin{equation}
\ddot{\phi}+3H\dot{\phi}+V'=0\label{field}
\end{equation}
\begin{equation}
H^2=\frac{\rho}{3m_P^2}\label{H}
\end{equation}
\begin{equation}
d(a^3\rho)=-p\,d(a^3)\label{emc}
\end{equation}
where \mbox{$H\equiv\dot{a}/a$} is the Hubble parameter with $a(t)$ being the 
scale factor of the Universe and the dot denotes derivative with respect to 
the cosmic time $t$. Also, $\rho$ and $p$ are the total energy and pressure 
density of the Universe respectively for which,

\begin{equation}
\rho=\rho_\phi+\rho_B
\hspace{1cm}
p=p_\phi+p_B
\end{equation}
where the subscript \mbox{\small $B$} denotes the background energy and 
pressure density which satisfy the equation of state,
\mbox{$p_B=w_B\rho_B$}, where $w_B=1/3$ \{0\} for the 
radiation \{matter\} era. Similarly,

\begin{equation}
w_\phi\equiv\frac{p_\phi}{\rho_\phi}=
\frac{\dot{\phi}^2-2V(\phi)}{\dot{\phi}^2+2V(\phi)}
\label{phi}
\end{equation}
The Universe is undergoing accelerated expansion when 
\mbox{$\rho_\phi>\rho_B$} and \mbox{$w_\phi<0$}.

\section{INFLATION}

During inflation (\ref{field}) and (\ref{H}) become,

\begin{eqnarray}
3H\dot{\phi}+V'=0 & \hspace{1cm} & 
H^2\simeq V/3 m_P^2
\label{infeqm}
\end{eqnarray}
The slow roll conditions are \mbox{$\varepsilon,\eta<1$} where, 

\begin{equation}
\varepsilon\equiv\frac{m_P}{\sqrt{6}}\frac{|V'|}{V}\hspace{1.5cm}
\eta\equiv\frac{m_P^2}{3}\frac{|V''|}{V}
\label{eps}
\end{equation}
When \mbox{$\phi=\phi_{\varepsilon}$} such that
\mbox{$\varepsilon(\phi_{\varepsilon})=1$} inflation is terminated. 
For \mbox{$\lambda\sim 1$} we have,
\mbox{$\phi_{\varepsilon}\approx 0.1\,\mu$}
Thus, during inflation \mbox{$\phi\ll\mu$} and, 

\begin{eqnarray}
V\simeq 2z^2\mu^4 = \mbox{const.} & &
V'\simeq -8z^3\,\frac{\mu^4}{m_P}
\label{saa}
\end{eqnarray}
Using these we find,

\begin{eqnarray}
N=\int_{\phi_{\varepsilon}}^{\phi_N}\frac{V}{V'}\frac{d\phi}{m_P^2}
& \Rightarrow & 
\phi_N\simeq\phi_{\varepsilon}-4zNm_P
\label{fN}
\end{eqnarray}
where \mbox{$N\equiv\int_{\varepsilon}^{^N}(da/a)$} is the remaining
number of e-foldings until the end of inflation. 

\section{INFLATIONARY REQUIREMENTS}

\subsection{Seeding LSS and CMBR}

The amplitude of the density perturbations is,

\begin{equation}
\frac{\delta\rho}{\rho}\;\simeq\;\frac{H^2}{\dot{\phi}}
\simeq -\frac{1}{\sqrt{3}}\frac{V^{3/2}}{m^3_PV'}\simeq
\frac{1}{2\sqrt{6}}(\frac{\mu}{m_P})^2
\end{equation}
COBE observations constrain the above as, 

\begin{equation}
\frac{\Delta T}{T}\simeq\frac{\delta\rho}{\rho}\simeq 10^{-5}
\;\Rightarrow\;
%(\frac{\mu}{m_P})^2\simeq 5\times 10^{-5}
\mu\simeq 1.7\times 10^{16}\mbox{GeV}
\label{mu}
\end{equation}
Thus, $\mu$ is given by the scale of grand unification.

Let us turn to the power spectrum of density perturbations. 
The spectral index is defined as, 

\begin{equation}
n_s\equiv 1+\frac{d\ln P_k}{d\ln k}
\end{equation}
where \mbox{$k^{-1}\sim He^N$} is the relevant scale at the end of inflation 
and, \mbox{$(P_k)^{1/2}\sim\frac{\delta\rho}{\rho}(k)$}. After some algebra 
one finds \cite{mine},

\begin{equation}
n_s-1\simeq 8\lambda^2-2\Big(4\omega^2-\frac{1}{2}\;\Big)\pi^2
\end{equation}
>From COBE, \mbox{$n_s\approx 1$}, which is ensured by taking,

\begin{equation}
2\sqrt{2}\,\lambda\approx\pi\sqrt{8\omega^2-1}
\label{lcond}
\end{equation}
Thus, we require \mbox{$\omega\geq\frac{1}{\sqrt{8}}$} so that,

\begin{equation}
\frac{1}{\sqrt{8}}\leq\omega<\frac{1}{2}
\label{range}
\end{equation}

\subsection{Horizon \& Flatness}

The Horizon and Flatness problems are solved if
the scale $H_0^{-1}$ of the horizon at present did exit the horizon 
during the inflationary period. The remaining e-foldings $N_H$ 
when this happened are,

\begin{equation}
N_H\simeq\ln(\frac{T_0}{H_0})+\ln(\frac{H_{\rm end}}{T_{\rm reh}})
\label{NH}
\end{equation}
where $T_{\rm reh}$ is the reheating temperature.
Using (\ref{fN}) we have the constraint,

\begin{equation}
z\leq\frac{\phi_\varepsilon}{4N_Hm_P}
\label{zcons}
\end{equation}

\section{AFTER THE END OF INFLATION}

\subsection{Reheating}

Although the inflaton does not decay into a thermal bath of SM particles, 
particle production is generated on gravitational grounds. 
The gravitationally generated fields 
have an approximately scale invariant spectrum with amplitude given by the 
Hawking temperature \cite{grav}\cite{gravkin}. Consequently, 

\begin{equation}
T_{\rm reh}\sim H_{\rm end}/2\pi
\label{Treh}
\end{equation}
In view of (\ref{NH}) this gives, \mbox{$N_H\simeq 69$}. Therefore, 
(\ref{zcons}) suggests that \mbox{$z\sim 10^{-6}$}. Such a small $z$ 
ensures the validity of the slow-roll conditions. Using (\ref{H}) we find, 
\mbox{$T_{\rm reh}\sim 10^9$GeV}. Thus, grand unification is not restored 
so that there is no monopole production. Moreover, such a low $T_{\rm reh}$
satisfies also the gravitino constraints.

\subsection{Kination}

One can write (\ref{field}) as, 

\begin{equation}
\frac{\dot{\rho}_{\rm kin}}{\dot{V}}=-1-3H\frac{\dot{\phi}}{V'}
\end{equation}
where \mbox{$\rho_{\rm kin}\equiv\frac{1}{2}\dot{\phi}^2$}.
This suggests, \mbox{$\dot{\rho}_{\rm kin}\gg\dot{V}$} during inflation. 
Since \mbox{$\varepsilon=1\Leftrightarrow\rho_{\rm kin}\simeq V_{\rm end}$},
after the end of inflation $\rho_{\rm kin}$ dominates over $V$.
Then \mbox{$\rho_\phi=p_\phi=\rho_{\rm kin}$}, which, using
(\ref{emc}), gives, \mbox{$\rho_\phi\propto a^{-6}$}. 

Gravitational reheating is not prompt. Indeed, 
\mbox{$\rho_B\sim T_{\rm reh}^4\sim H_{\rm end}^4\ll 
V_{\rm end}\sim\rho_{\rm kin}$}. 
Thus, the Universe remains $\phi$-dominated. This period of 
$\rho_{\rm kin}$ domination is usually called kination or deflation
\cite{quinfkin}\cite{gravkin}.
Because \mbox{$\rho_B\propto a^{-4}$}, radiation
will eventually dominate as required by the Hot Big Bang.

Since \mbox{$\rho\propto t^{-2}$} during kination, \mbox{$a\propto t^{1/3}$}. 
Using, (\ref{infeqm}) and (\ref{Treh}) we find that kination ends at,

\begin{equation}
t_*\sim t_{\rm end} (\frac{m_P^4}{V_{\rm end}})^{3/2}
\label{t*}
\end{equation}
where \mbox{$t_{\rm end}\sim m_P/\sqrt{V_{\rm end}}$} is the time when 
inflation is terminated. In view of (\ref{saa}) we find 
\mbox{$t_*\sim 10^{-2}$sec}, i.e. earlier than Nucleosynthesis.

Let us find how much does the scalar field roll during kination. 
In this regime (\ref{field}) becomes,
\mbox{$\ddot{\phi}+(\dot{\phi}/t)\simeq 0$}.
Solving we find,

\begin{equation}
\phi_*\equiv\phi(t_*)\simeq\phi_\varepsilon+\frac{\sqrt{3}}{4\pi}
\ln(\frac{m_P}{V^{1/4}_{\rm end}})\,m_P
\label{f*}
\end{equation}
%where \mbox{$V_{\rm end}^{1/4}\sim 10^{13}$GeV}.

\subsection{Hot Big Bang}

At the onset of the radiation era $\rho_{\rm kin}$
decreases rapidly as, \mbox{$\rho_{\rm kin}\propto t^{-3}$} 
since \mbox{$a\propto t^{1/2}$}. Thus, in little time 
\mbox{$\dot{\phi}\rightarrow 0$} and the field freezes at $\phi_F$. Indeed, 
it can be shown \cite{mine} that,

\begin{equation}
\phi_F\!=\!\phi_*\!+2\dot{\phi}_*t_*\!
\simeq\phi_\varepsilon+\frac{\sqrt{3}}{12\pi}
\Big[1+3\ln(\frac{m_P}{V_{\rm end}^{1/4}})\Big]m_P
\label{ffgr}
\end{equation}
which evaluates to, \mbox{$\phi_F\simeq 1.68\,m_P\gg\mu$}. 

\section{QUINTESSENCE}

When \mbox{$\phi\gg\mu$} then the potential becomes,

\begin{equation}
V(\phi)\simeq\frac{\pi^2}{8}\,\frac{M^8}{\phi^4}
\left[1+\frac{\sqrt{z}}{\cos(\omega\pi)}(\frac{\mu}{M})^2\,
e^{-\lambda\phi/\mu}\right]^4
\label{quint}
\end{equation}
The exponential term in (\ref{quint}) dominates when,

\begin{equation}
\phi<\phi_C\equiv\frac{\mu}{\lambda}
\Big\{2\ln(\frac{\mu}{M})-\ln[\cos(\omega\pi)]
+\frac{1}{2}\ln z\Big\}
\label{fc}
\end{equation}
The quintessential part of the evolution of $\phi$
can be explored analytically with the use of (\ref{field}),

\begin{equation}
3H\dot{\phi}\simeq -V'(\phi)\Rightarrow
-\!\!
\int_{_{\mbox{\scriptsize $\phi_{\!_F}$}}}^{^{^{\mbox{\scriptsize 
$\phi\mbox{\tiny (}t\mbox{\tiny )}$}}}}
\hspace{-0.4cm}\frac{d\phi}{V'(\phi)}=\frac{1}{4}(w_B+1)t^2
\label{atrfe}
\end{equation}
where the field's acceleration $\ddot{\phi}$ 
is ignored because $\phi$
lies at the ``quintessential tail'' of $V(\phi)$. 
It is found \cite{mine} that either $\phi$ remains frozen 
at $\phi_F$ or it follows attractor solutions of the form
\mbox{$\phi=\phi_{\rm atr}(t)$}, so that 
\mbox{$V(\phi)=$ min$\{V(\phi_F),V(\phi_{\rm atr})\}$}.

\section{QUINTESSENCE REQUIREMENTS}

The coincidence problem translates to,

\begin{equation}
V(\phi_0)=\Omega_\phi\rho_0
\label{coin}
\end{equation}
where \mbox{$\phi_0\equiv\phi(t_0)$} is evaluated at the present time $t_0$ 
and \mbox{$\Omega_\phi\simeq 0.65$} is the ratio of $\rho_\phi$ over the 
critical density today \cite{dark} \mbox{$\rho_0\simeq 10^{-120}m_P^4$}.

Moreover, in order to explain the current accelerated expansion of the 
Universe we require,

\begin{equation}
w_\phi=\frac{\varepsilon_0^2-
\Omega_\phi^{-1}}{\varepsilon_0^2+\Omega_\phi^{-1}}<0
\label{wf}
\end{equation}
where \mbox{$\varepsilon_0\equiv\varepsilon(\phi_0)$} and we used
(\ref{phi}) and (\ref{eps}).

It can be shown that, if $\phi$ starts following the attractor solutions 
it is impossible to meet the above requirements in the available parameter 
space \cite{mine}. Thus, we need \mbox{$\phi_0=\phi_F$}, which requires,

\begin{equation}
V(\phi_0)\leq V(\phi_{\rm atr}(t_0))
\label{frozen}
\end{equation}

\subsection{Regimes}

\medskip

{\boldmath\bf $\phi<\phi_C$: Quasi-exponential quintessence}

\medskip

The potential reduces to,

\begin{equation}
V(\phi)\simeq\frac{(\pi z)^2/8}{[\cos(\omega\pi)]^4}
\,\frac{\mu^8}{\phi^4}\,e^{-4\lambda\phi/\mu}
\label{exp}
\end{equation}
Using (\ref{mu}) we find from (\ref{coin}) that, 
\mbox{$\lambda\simeq 0.224$}. Moreover, \mbox{$\phi_{F}\leq\phi_C$} 
results in, \mbox{$M\leq 0.82$ TeV}. These values can be shown \cite{mine} 
to satisfy the constraint (\ref{frozen}). Finally, (\ref{wf}) gives, 
\mbox{$w_\phi\simeq -0.836$}. 

\medskip

{\boldmath\bf $\phi>\phi_C$: Inverse power-law quintessence}

\medskip

The potential reduces to,

\begin{equation}
V(\phi)\simeq \frac{\pi^2}{8}\,\frac{M^8}{\phi^4} 
\label{invpl}
\end{equation}
Now (\ref{coin}) gives, 
\mbox{$M\simeq 2.92$ TeV}. Using this, (\ref{frozen}) gives,
\mbox{$\lambda\geq 0.225$}, corresponding to practically all the 
$\omega$-range (\ref{range}). These values, satisfy 
\mbox{$\phi_{F}\geq\phi_C$} as required \cite{mine}. Finally, (\ref{wf}) 
gives, \mbox{$w_\phi\simeq -0.228$}.

\section{CONCLUSIONS}

We have studied analytically a model of quintessential inflation and showed 
that it is able to satisfy all the requirements of inflation and quintessence
with natural values of the parameters, i.e. \mbox{$M\sim$ TeV}, which 
is possible to link with supersymmetry breaking or even large extra 
dimensions, and \mbox{$\mu\sim 10^{16}$GeV}, which is the grand unification 
scale. The rest of the parameters also assume natural values, 
\mbox{$\lambda,\omega\lsim 1$} except \mbox{$z\sim 10^{-6}$}, which 
corresponds to the typical fine tuning for the flatness of inflationary 
potentials. Even though the theoretical justification of our model is still 
an open issue we have shown that it is indeed possible to construct successful 
quintessential inflationary models without the need of any additional 
fine-tuning for quintessence. In that sense, our model outshines the 
cosmological constant alternative.

\end{document}